\documentclass[preprint,superscriptaddress,amsmath,amssymb,aps,prb]{revtex4-1}
\usepackage{graphicx}
\usepackage{bm}
\usepackage{amssymb}
\usepackage{color}
\usepackage{amsmath}
\usepackage[colorlinks=true,linkcolor=blue,anchorcolor=black,citecolor=blue,filecolor=black,menucolor=black,runcolor=black,urlcolor=blue]{hyperref}

\begin{document}
\title{Flatbands from Bound States in the Continuum for Orbital Angular Momentum Localization}

\affiliation{College of Physics and Optoelectronic Engineering, Ocean University of China, Qingdao 266100, China}
\affiliation{Research Center of Fluid Machinery Engineering and Technology, School of Physics and Electronic Engineering, Jiangsu University, Zhenjiang 212013, China}
\author{Weiwei Zhu}
\thanks{These authors have contributed equally to this work.}
\affiliation{College of Physics and Optoelectronic Engineering, Ocean University of China, Qingdao 266100, China}
\affiliation{Key Laboratory for Optics Photoelectronics, Qingdao 266100, China}
\affiliation{Engineering Research Center of Advanced Marine Physical Instruments and Equipment of MOE, Qingdao 266100, China}
\author{Hongyu Zou}
\thanks{These authors have contributed equally to this work.}
\affiliation{Research Center of Fluid Machinery Engineering and Technology, School of Physics and Electronic Engineering, Jiangsu University, Zhenjiang 212013, China}
\author{Yong Ge}
\affiliation{Research Center of Fluid Machinery Engineering and Technology, School of Physics and Electronic Engineering, Jiangsu University, Zhenjiang 212013, China}
\author{Yin Wang}
\affiliation{Research Center of Fluid Machinery Engineering and Technology, School of Physics and Electronic Engineering, Jiangsu University, Zhenjiang 212013, China}
\author{Zheyu Cheng}
\affiliation{Division of Physics and Applied Physics, School of Physical and Mathematical Sciences, Nanyang Technological University, Singapore 637371, Singapore}
\author{Bing-bing Wang}
\affiliation{Research Center of Fluid Machinery Engineering and Technology, School of Physics and Electronic Engineering, Jiangsu University, Zhenjiang 212013, China}
\author{Shou-qi Yuan}
\affiliation{Research Center of Fluid Machinery Engineering and Technology, School of Physics and Electronic Engineering, Jiangsu University, Zhenjiang 212013, China}
\author{Hong-xiang Sun}
\email{jsdxshx@ujs.edu.cn}
\affiliation{Research Center of Fluid Machinery Engineering and Technology, School of Physics and Electronic Engineering, Jiangsu University, Zhenjiang 212013, China}
\affiliation{State Key Laboratory of Acoustics, Institute of Acoustics, Chinese Academy of Sciences, Beijing 100190, China}
\author{Haoran Xue}
\email{haoranxue@cuhk.edu.hk}
\affiliation{Department of Physics, The Chinese University of Hong Kong, Shatin, Hong Kong SAR, China}
\author{Baile Zhang}
\email{blzhang@ntu.edu.sg}
\affiliation{Division of Physics and Applied Physics, School of Physical and Mathematical Sciences, Nanyang Technological University, Singapore 637371, Singapore}
\affiliation{Centre for Disruptive Photonic Technologies, Nanyang Technological University, Singapore, 637371, Singapore}

\maketitle

\textbf{A flatband material is a system characterized by energy bands with zero dispersion, allowing for the compact localization of wavefunctions in real space. This compact localization significantly enhances inter-particle correlations and light-matter interactions, leading to notable advancements such as fractional Chern insulators in condensed matter systems and flat-band lasers in photonics. Previous flatband platforms, including twisted bilayer graphene and artificial kagome/Lieb lattices, typically focused on nondegenerate flatbands, lacking access to the high degeneracy that can facilitate the localization of orbital angular momentum (OAM). Here, we propose a general framework to construct highly degenerate flatbands from bound states in the continuum (BICs)--a concept originating from quantum theory but significantly developed in photonics and acoustics in recent years. The degeneracy of flatbands is determined by the number of BICs within each unit cell in a lattice. We experimentally validate this approach in two-dimensional (2D) and three-dimensional (3D) acoustic crystals, demonstrating flatbands with 4-fold and 12-fold degeneracies, respectively. The high degeneracy provides sufficient internal degrees of freedom, enabling the selective excitation of localized OAM at any position in any direction. Our results pave the way for exploring BIC-constructed flatbands and their localization properties. }

Flatband materials with dispersionless energy band feature compact localized bulk states~\cite{PhysRevB.34.5208,PhysRevB.54.R17296,PhysRevB.78.125104,PhysRevB.95.115135,Leykam2018}. Different from the usual localized states that exist in systems with defects~\cite{PhysRevLett.58.2486,PhysRevLett.58.2059,science284,Qi2004,Rinne2008} or disorders~\cite{PhysRev.109.1492,segev2013}, here the compact localized states exist in the bulk of perfectly periodic lattices. Their localization mechanism can be understood by destructive interference, as examples shown in different platforms, such as twisted bilayer graphene~\cite{pnas1108174108,PhysRevLett.126.223601}, kagome lattice~\cite{Mielke_1991a,Mielke_1991b} and Lieb lattice~\cite{PhysRevLett.62.1201}. The compact localization accompanied with high density of states significantly enhances inter-particle correlations and light-matter interactions, leading to strong correlated states like fractional Chern insulators~\cite{PhysRevLett.106.236802,PhysRevLett.106.236803,PhysRevLett.106.236804} and having potential applications in distortion-free storage~\cite{PhysRevLett.113.236403,PhysRevB.98.134203,PhysRevB.102.121302}, quantum state transfer~\cite{PhysRevA.96.043803,PhysRevLett.123.080504} and reconfigurable laser arrays~\cite{mao2021,luan2023}.

Up to now, flatband materials have been experimentally realized in diverse context, including condensed matter physics~\cite{Cao2018,sciadvaau4511,Wakefield2023}, photonics~\cite{PhysRevLett.114.245504,PhysRevLett.116.183902,Cantillano_2018,Wang2020} and phononics~\cite{Xue2019,PhysRevLett.132.266602,Oudich2024}. However, most works focused on nondegenerate flatbands, lacking access to the high degeneracy with enhanced density of states and internal degrees of freedom, which would be beneficial to the aforementioned flatband effects and can facilitate the localization of orbital angular momentum (OAM). Only a few theoretical works have studied how to construct multiple degenerate flatbands using case studies~\cite{PhysRevA.100.043841,PhysRevB.102.235117,Nakai2022,PhysRevB.109.035140}. OAM, as a degree of freedom other than amplitude, polarization, and phase of structured waves, has revolutionized applications in optical communications~\cite{yao2011orbital,science1237861} and quantum information processing~\cite{Mair2001,science1190523}. However, there is no work discussing how to construct flatband modes with OAM or compact localization of OAM, which may be used to realize reconfigurable OAM laser arrays. A general method to design compact localization of OAM at any position in any direction is still lacking.

\begin{figure*}[htp]
\includegraphics[width=\linewidth]{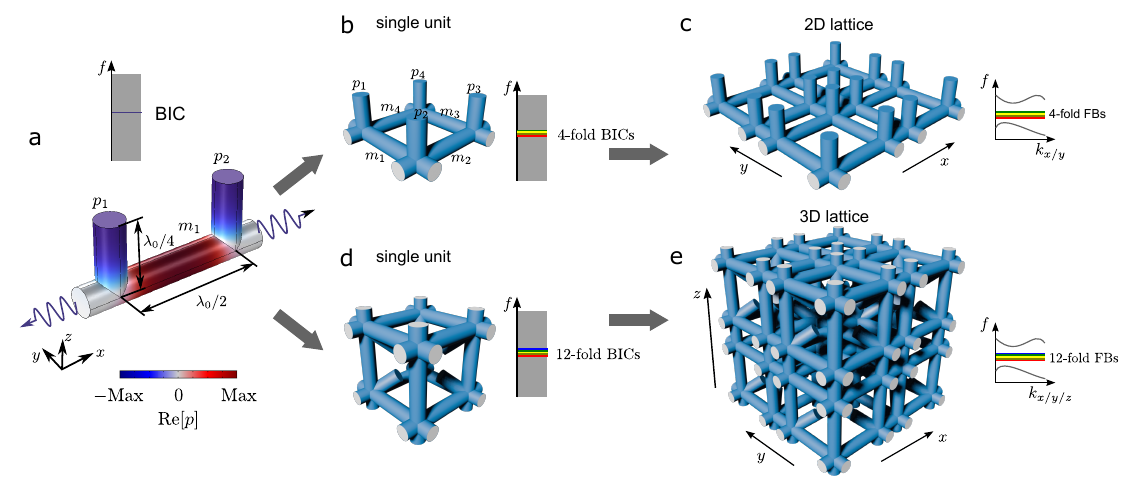}
\caption{\textbf{Degenerate BIC and multiple-fold flatband.} \textbf{a}, An open system that supports BIC. The system is composed of two resonant waveguides along $z$ and one transport waveguide along $x$. The length of resonant waveguides is $\lambda_0/4$ and the distance between nearest neighbor resonant waveguides is $\lambda_0/2$. Radiation boundary condition is applied in $x$ direction. \textbf{b} and \textbf{d}, 2D and 3D open system that supports degenerate BICs. \textbf{c} and \textbf{e}, 2D lattice and 3D lattice made up of 2D and 3D system in \textbf{b} and \textbf{d}. They support 2D 4-fold flatbands and 3D 12-fold flatbands, respectively.}
\label{fig1}
\end{figure*}

In this work, we propose a general framework for designing multifold flatbands for arbitrary dimensional systems based on bound states in the continuum (BICs), which are non-radiative modes that live within the continuous spectrum~\cite{Hsu2016}. The BICs are transferred to flatbands upon arranging the unit structure hosting BICs into a lattice. The unit structure with BICs can be any type of open wave system, and the lattice arrangement can also be in an arbitrary space group, regardless of the lattice dimension. Such generality allows us to realize flatbands with high degeneracy in high dimensions. We present two concrete models of acoustic crystals in two and three dimensions, which support 4-fold and 12-fold flatbands, respectively. We further show the multifold flatbands can support compact localization and reconfigurable tuning of OAM, which is inaccessible in nondegenerate flatbands. These flatbands and the compact localization of OAM are validated through both experiments and numerical simulations. We also present an electromagnetic design in Supplementary Information.

\textit{General construction.}----Our starting point is an open single unit supporting a few radiative modes and one or several non-radiative modes (i.e., BICs). Thanks to the open nature of the structure, we can easily construct a lattice by periodically arranging the single unit and connecting each unit to its neighbors through the open parts. After this process, the original radiative modes in each unit will hybridize with radiative modes in other units, forming dispersive Bloch bands. By contrast, the BICs, owing to their non-radiative characteristics, will stay trapped in one unit cell and will not couple with other modes, either in the same or neighboring unit cells. Thus, the BICs will be transformed into flatbands. Through such a straightforward picture we can see that the properties of the flatbands designed by this approach are totally determined by those of the single unit and the associated BICs. This greatly simplifies the construction of flatband systems, especially for those in higher dimensions and with high degeneracy.

The geometry of the single unit determines the possible connectivity when constructing the lattice and thus controls the symmetry and dimension of the resulting system. As we show later, by designing the geometry of the single unit, we can easily realize flatband systems in different dimensions. Furthermore,  as all BICs in the single unit will become flatband states, the detailed mode profiles and degeneracy of BICs and flatbands will be exactly the same. This elegant correspondence allows us to have full and detailed control over flatbands, not just their existence.

{\it Waveguide network model.--}
Based on the above scheme, we now introduce a waveguide network model to realize various flatband lattices. Consider the single unit shown in Fig.~\ref{fig1}\textbf{a}, which is composed of two resonant waveguides and one transport waveguide. Open boundary conditions are introduced to the two ends of the transport waveguide to allow radiation into free space. The resonant waveguides play the role of resonators with resonant frequency $v/\lambda_0$, where $v$ is the wave speed and $\lambda_0$ is the wavelength in free space. The two resonant waveguides can interact with each other by the propagating field in the transport waveguide.

\begin{figure*}
	\includegraphics[width=\linewidth]{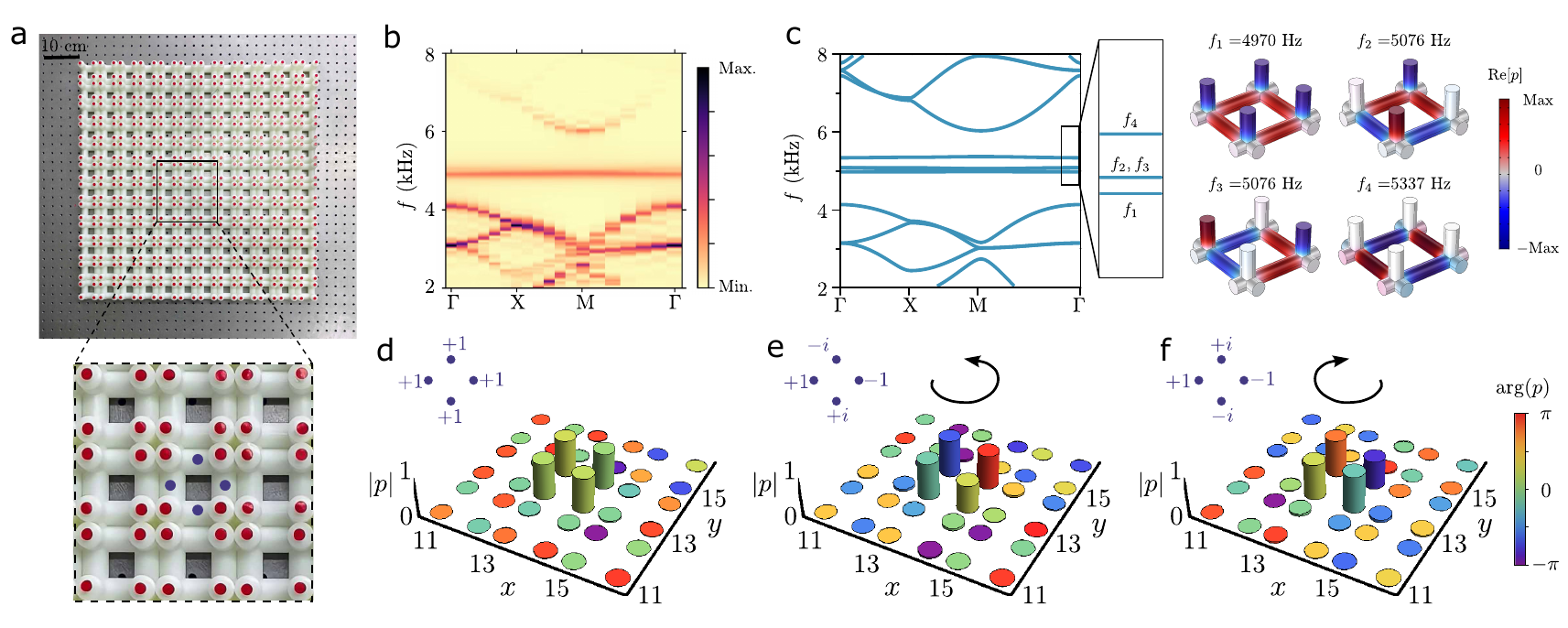}
	\caption{ \textbf{Four-fold flatbands for 2D lattice and compact localization of OAM}. \textbf{a}, Photograph of a sample with $12\times 12$ unit cells. The lower panel shows a zoom-in view of the sample. The source is marked by the purple dots. \textbf{b}, Measured band structure for 2D lattice in \textbf{a}. \textbf{c}, Simulated band structure. The right panel shows the simulated eigenfield for the four flatband modes at $\Gamma$ point. \textbf{d}, Measured localized flatband modes. \textbf{e}, Similar to \textbf{d} but with clockwise OAM. \textbf{f}, Similar to \textbf{d} but with anticlockwise OAM. }
	\label{fig2}
\end{figure*}

Due to the open boundaries, the eigenmodes of this structure are generally lossy, with BICs being the  exceptions~\cite{PhysRevLett.121.124501,Huang2023}. Specifically, a BIC can be obtained at the resonant frequency $v/\lambda_0$ by setting the distance between the two resonant waveguides as $\lambda_0/2$. The eigenmode property can be obtained by the following theoretical analysis. At the resonant frequency $v/\lambda_0$, the field distribution in the resonant and transport waveguides is standing-wave-like to make the amplitude at the intersections fix to zero. The eigenmode can be obtained by a superposition state $|\psi_{\mathrm{BIC}}\rangle=p_1|p_1\rangle+p_2|p_2\rangle+m_1|m_1\rangle$, where $|p_1\rangle$ and $|p_2\rangle$ ($|m_1\rangle$) are the wave shape in the resonant waveguides (transport waveguide between two resonant waveguide), $p_1$ and $p_2$ ($m_1$) are the corresponding amplitudes. These amplitudes need to meet the velocity conservation at the two intersections~\cite{PhysRevLett.118.084302}, which are $p_1+m_1=0$ and $p_2+m_1=0$. The eigenmode then become $|\psi_{\mathrm{BIC}}\rangle=m_1(-|p_1\rangle-|p_2\rangle+|m_1\rangle)$, which is exactly a BIC mode. Next, if we use this single unit as a unit cell to build a 1D lattice, we can get a flatband at frequency $v/\lambda_0$.

The dimension and degeneracy can be further increased by changing the single unit design. To show this, consider the structure displayed in Fig.~\ref{fig1}\textbf{b}, which can be obtained by performing a four-fold rotation operation on the previous structure. Similarly, the eigenmode at the resonant frequency $v/\lambda_0$ can be constructed by $|\psi_{\mathrm{4BIC}}\rangle=p_1|p_1\rangle+p_2|p_2\rangle+p_3|p_3\rangle+p_4|p_4\rangle+m_1|m_1\rangle+m_2|m_2\rangle+m_3|m_3\rangle+m_4|m_4\rangle$, which contains eight degrees of freedom.
Considering the velocity flow conservation at four interactions, the eight coefficients meet $p_1+m_1+m_4=0$, $p_2+m_1+m_2=0$, $p_3+m_2+m_3=0$ and $p_4+m_3+m_4=0$. Only four undetermined coefficients are left.
We choose $m_1$, $m_2$, $m_3$ and $m_4$ to represent the eigenfunction.
The eigenfunction becomes $|\psi_{\mathrm{4BIC}}\rangle=m_1(-|p_1\rangle-|p_2\rangle+|m_1\rangle)+m_2(-|p_2\rangle-|p_3\rangle+|m_2\rangle)+m_3(-|p_3\rangle-|p_4\rangle+|m_3\rangle)+m_4(-|p_4\rangle-|p_1\rangle+|m_4\rangle)$, which is a superposition of four linearly independent eigenstates and each of them is a BIC~\cite{PhysRevLett.121.124501,Huang2023}). Therefore, this single unit supports 4-fold degenerate BICs at frequency $v/\lambda_0$. Moreover, according to the geometry, this single unit can be arranged into a 2D lattice (Fig.~\ref{fig1}\textbf{c}), with a 4-fold flatband. Later, we will also show such 4-fold degenerate BICs can be used to realize compact localization of OAM.

More generally, the degeneracy of eigenfunction can be captured by $N=N_w-N_n$, where $N_w$ is the number of waveguides and $N_n$ is the number of nodes. For the 2D structure, we have $N_w=8$ and $N_n=4$. In a similar manner, we can also design a single unit with other values of $N_w$ and $N_n$ and with the capability to build 3D lattices. As an extreme case, we show in Fig.~\ref{fig1}\textbf{d} a design for a 12-fold BIC containing 12 transport waveguides, 8 resonant waveguides and 8 nodes, which can be used to obtain a 12-fold flatband in a 3D lattice (Fig.~\ref{fig1}\textbf{e}). We note that the lattice configurations are not limited to the square and cubic lattices shown in Fig.~\ref{fig1} but can also be of other types like the triangle lattice (see Supplementary Information).

\begin{figure*}[htp]
\includegraphics[width=0.9\linewidth]{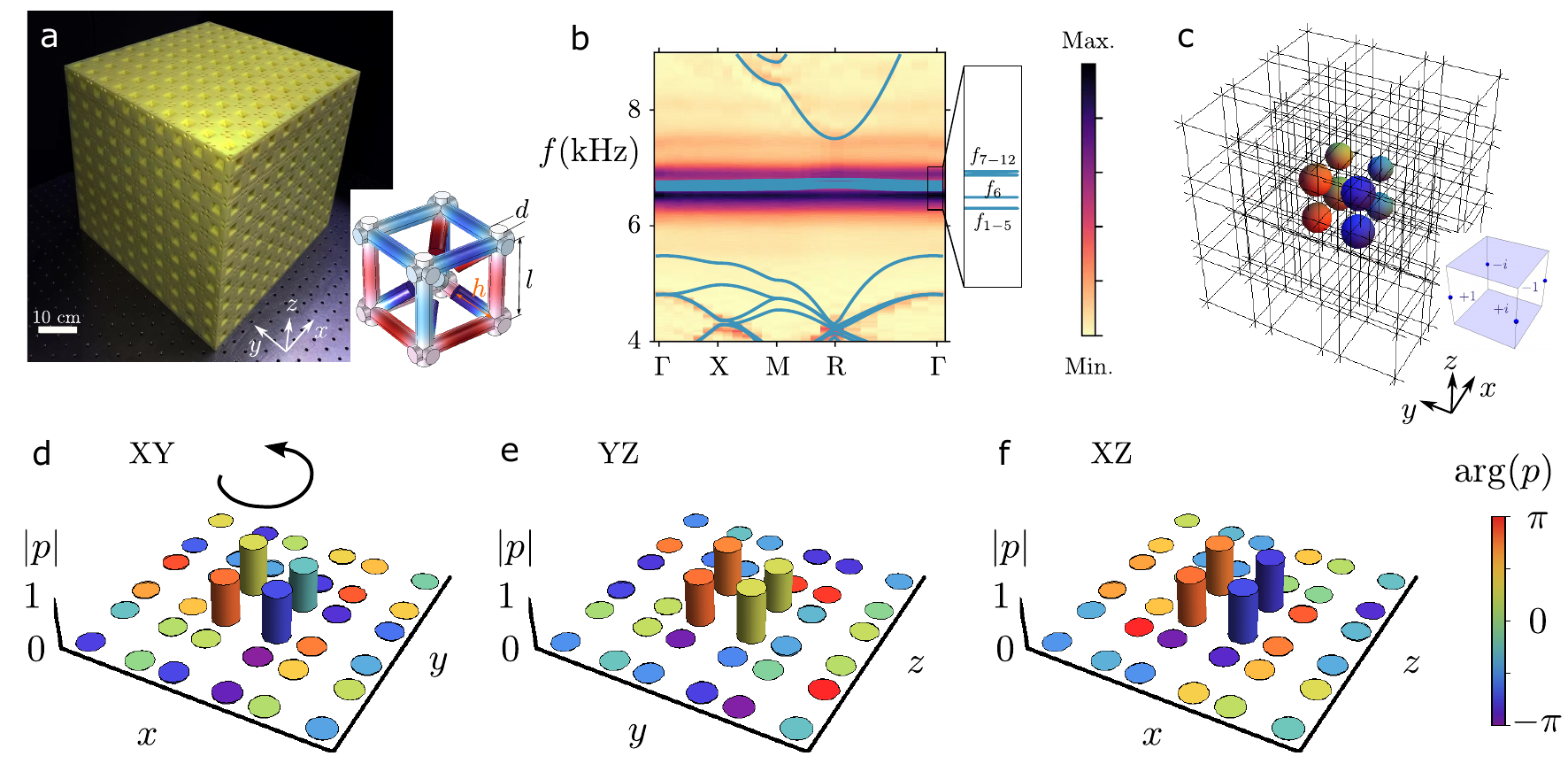}
\caption{\textbf{Twelve-fold flatband for 3D lattice and compact localization of OAM}. \textbf{a}, Photograph of a sample with 10$\times$10$\times$10 unit cells. The lower-right panel shows the mode profile of a flatband mode. \textbf{b}, Measured and simulated band structures along the high symmetric line in the Brillouin zone. The simulated band structure is shown in the blue lines. \textbf{c}, Measured field distributions at the flatband frequency. The lower-right panel shows the locations of the four sources. \textbf{d}-\textbf{f}, Cross-section plots of the field distribution in \textbf{c}. \textbf{d} for the middle layer of the XY plane, \textbf{e} for the middle layer of the YZ plane and \textbf{f} for the middle layer of the XZ plane.}
\label{fig3}
\end{figure*}

{\it Experimental realization of four-fold flatbands in a 2D acoustic crystal.--}
Next we consider a realistic model of constructing BIC. The model is still composed of two resonant waveguides and one transport waveguide as in Fig.~\ref{fig1}\textbf{a}. The only difference is here we take into account the diameter of the waveguide. We fix the diameter of all waveguides and the height of the resonant waveguides as $d=8$ mm and $h=20$ mm, respectively, while leaving the distance between two resonant waveguides $l$ as a tunable parameter. According to the theoretical prediction, BIC occurs at $l=2h$ for a single open unit. However, due to the nonzero cross-section area of waveguides in the realistic design, this condition will be slightly shifted. To accurately determine the BIC condition, we compute the eigenfrequencies for different values of $l$ using a finite-element method (see Method for numerical details). A BIC with a pure real eigenfrequency is found when $l=36$ mm . It is worth noting that the BIC's frequency is around 5 kHz, which differs from the theoretical prediction $v/(4h)\approx 4.288$ kHz due to the fact that the diameter of the waveguides is comparable with $h$.

The 1D acoustic waveguide system can be used as building blocks to construct composite structures in high dimensional systems, which support degenerate BICs and can be used as unit cells to construct an acoustic crystal supporting multifold flatbands. Figure~\ref{fig2}\textbf{a} shows a sample of a 2D acoustic crystal on a square lattice. The whole structure is filled with air and bounded by photosensitive resins that act as rigid walls for sound. Here, each unit cell contains four resonators and four transport waveguides, with lattice constant $a_s=54$ mm. As predicted by our theory, the unit cell supports 4-fold degenerate BICs and the corresponding 2D lattice supports 4-fold degenerate flatbands. 

We measure the acoustic field pressure excited by a single source placed at the center of the sample (see Methods for more experimental details). By a Fourier transform, we obtain the band structure of the system. The measured and numerically simulated dispersions are plotted in Fig.~\ref{fig2}\textbf{b} and \ref{fig2}\textbf{c}, respectively, both of which clearly exhibit flatbands around 5 kHz. In the right panel of Fig.~\ref{fig2}\textbf{c}, we plot the eigenfields of the four flat band modes at $\Gamma$ point. Due to the nonzero cross-section area of the acoustic waveguides, the frequencies of the flatbands slightly split. One eigenmode has a frequency larger than others. Note that since the experimental signal is measured from the resonant waveguides, the flatband with a higher frequency than the rest three is not observed in the experiment due to its neglectable field distribution in the resonant waveguides (see the eigen profiles in Fig.~\ref{fig2}\textbf{c}).

{\it Compact localization of OAM.--}
While compact localized states can already be achieved from a nondegenerate flatband, one advantage of multifold flatbands is that the localization profile is highly tunable thanks to the high degeneracy.  As our theory predicts, the eigenfunction of localized states in the 2D model is $|\psi_{\mathrm{4BIC}}\rangle=m_1(-|p_1\rangle-|p_2\rangle+|m_1\rangle)+m_2(-|p_2\rangle-|p_3\rangle+|m_2\rangle)+m_3(-|p_3\rangle-|p_4\rangle+|m_3\rangle)+m_4(-|p_4\rangle-|p_1\rangle+|m_4\rangle)$.
We can choose different $m_1$, $m_2$, $m_3$ and $m_4$ to obtain different localized states. For example, if we choose $m_1=m_2=m_3=m_4=+1$, we obtain $|\psi_{\mathrm{4BIC}}\rangle=-2|p_1\rangle-2|p_2\rangle-2|p_3\rangle-2|p_4\rangle+|m_1\rangle+|m_2\rangle+|m_3\rangle+|m_4\rangle$. Such a state is observed in experiment by four excitations in the middle of transport waveguides. The results are shown in Fig.~\ref{fig2}\textbf{d}, where the relative amplitude and phase are the same in the four resonators. Interestingly, if we choose $m_1=+1$, $m_2=+i$, $m_3=-1$ and $m_4=-i$, we realize compact localization of OAM with anticlockwise angular momentum, $|\psi_d\rangle=(-1+i)|p_1\rangle+(-1-i)|p_2\rangle+(1-i)|p_3\rangle+(1+i)|p_4\rangle+|m_1\rangle+i|m_2\rangle-|m_3\rangle-i|m_4\rangle$. The experimental results for such a state is shown in Fig.~\ref{fig2}\textbf{e}, where we can see the relative amplitude for four resonators are the same and relative phase increase $2\pi$ in anticlockwise direction. If we change the excitation to $m_1=+1$, $m_2=-i$, $m_3=-1$ and $m_4=+i$, we realize compact localization of OAM with clockwise angular momentum, $|\psi_d\rangle=(-1-i)|p_1\rangle+(-1+i)|p_2\rangle+(1+i)|p_3\rangle+(1-i)|p_4\rangle+|m_1\rangle-i|m_2\rangle-|m_3\rangle+i|m_4\rangle$. The experimental results for such a state is shown in Fig.~\ref{fig2}\textbf{f}, where we can see the relative amplitude for four resonators are the same and relative phase increase $2\pi$ in clockwise direction. Both theory and experimental results are consistent with simulation results (see Supplementary Information).

{\it Experimental realization of twelve-fold flatbands in a 3D acoustic crystal--}
Finally, we demonstrate an extreme case, 12-fold flatbands in three dimensions (Fig.~\ref{fig1}\textbf{e}), using an acoustic crystal. A photograph of the sample with 10$\times$10$\times$10 unit cells is displayed in Fig.~\ref{fig3}\textbf{a}. The unit cell's structure is similar to the theoretical model in the lower-right panel, with the height of resonant waveguide $h=17.2$ mm,  the diameter of waveguides $d=6$ mm and the lattice constant $a_s=40$ mm. Right-down panel of Fig.~\ref{fig3}\textbf{a} shows one of the twelve flatband states, which is compactly localized within the unit cell. We put one source in the sample and measure the field in the transport waveguides. Band structure can be obtained by Fourier transformation. The measured and numerically simulated band structures are given in Fig.~\ref{fig3}\textbf{b}. Flatbands emerge at around 6.95 kHz in both plots. From the numerically simulated dispersions, we further confirm the degeneracy is 12-fold, consistent with our theory. Such an extremely high degeneracy makes the Fourier intensity of the flatbands notably higher compared to the 1D and 2D cases.

Besides, we can use the high degeneracy properties to construct compact localization of OAM at any location in any direction. Due to the higher degeneracy of our system, we can obtain nonzero OAM along different directions by choosing proper excitation, $e.g.$ (1,0,0), (0,1,0), (0,0,1) or (1,1,1) direction. To excite the localized flatband modes with OAM pattern, we use four sources. The position and relative phases are shown in the lower-right panel of Fig.~\ref{fig3}\textbf{c}. Such four-sources excitation can realize compact localization of OAM in $z$ direction. The measured field amplitude and phase surrounding sources are shown in Fig.~\ref{fig3}\textbf{c}, where we notice most fields are localized in the central unit cell. To show more details, we plot the field of different cross-sections of Fig.~\ref{fig3}\textbf{c} in Figs.~\ref{fig3}\textbf{d}-\ref{fig3}\textbf{f}. Fig.~\ref{fig3}\textbf{d} shows results for the middle layer of $xy$ cross-section, where we find the field is mainly localized in the central unit cells and the phase of four central resonators change $2\pi$ in the anticlockwise direction, indicating a nonzero OAM in $z$ direction. From Figs.~\ref{fig3}\textbf{e} and \ref{fig3}\textbf{f}, we can confirm the mode is localized in all directions. It is worth mentioning that here we only excite a $z-$direction OAM mode, those in other directions can be realized by simply rotating the sources.

\begin{figure}[htp]
	\includegraphics[width=0.5\linewidth]{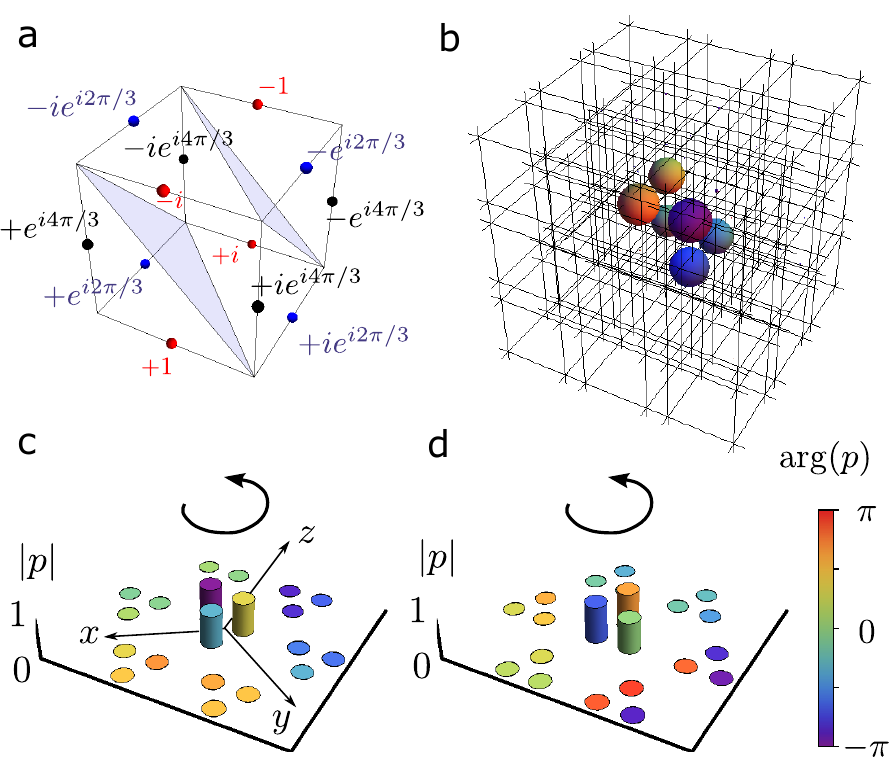}
	\caption{\textbf{Experimental results for compact localization of OAM in (1,1,1) direction}. \textbf{a}, The position and phase of twelve sources used to excite localized flat-band modes with OAM in (1,1,1) direction. \textbf{b}, The amplitude and phase of localized states in 3D space. \textbf{c} and \textbf{d}, Same as \textbf{b} but in cross section. \textbf{c} for upper layer of (1,1,1) surface and \textbf{d} for lower layer.}
	\label{fig4}
\end{figure}

We even can realize compact localization of OAM in (1,1,1) direction. Figure~\ref{fig4}\textbf{a} illustrates the twelve sources used to excite the flatband modes. The four red (blue/black) ones are used to excite OAM in $x$ ($y$/$z$) direction. As a whole, they excite the compact localized states with OAM in (1,1,1) direction. Fig.~\ref{fig4}\textbf{b} shows the measured field amplitude and phase surrounding the sources. We notice all the fields are localized in the center unit cells and mainly distributed in six resonators. The six resonators can be classified into two different (1,1,1) surfaces. Figs.~\ref{fig4}\textbf{c} and \ref{fig4}\textbf{d} show their amplitudes and phases, from which we see $2\pi$ phase change in anticlockwise direction for each (1,1,1) layer, a signature of nonzero OAM in (1,1,1) direction.

{\it Conclusion--} In summary, we have proposed a method to construct multifold flatbands based on degenerate BICs. Both theory, simulations and experiments demonstrate that the method works well. The high degeneracy allows compact localization of OAM at any position in any direction. These flatband modes would be useful for designing large-scale tunable OAM laser arrays. The model we designed only contains a class of waveguides. In our work, we show it can be realized in 2D and 3D airborne acoustic systems. Such a method can be extended to other systems, like transmission line networks~\cite{PhysRevLett.81.5540,Jiang2019} and superconducting circuits~\cite{PhysRevLett.83.5102,Kollar2019}.

{\it Methods--}

\textbf{Numerical simulations.} All simulations are performed using the acoustic module of COMSOL Multiphysics, which is based on the finite element method. The photosensitive resin used for sample fabrication is set as the hard boundary due to its large impedance mismatch with air. The real sound speed at room temperature is $c_0$ = 343 m/s. The density of air $\rho$ is set to be 1.28 kg/m$^3$. 

\textbf{Experimental measurements.} In the experiment, a broadband sound signal (2 Hz to 10 kHz) is used as a point-like sound source to excite the sound wave. The pressure of each site is detected by a microphone (Bruel\&Kjær Type 4961) adhered to a long tube (of diameter about 0.35 cm and a length of 35 cm). The signal is recorded and frequency-resolved by a multi-analyzer system (Bruel\&Kjær 3160-A-022 module). 

\textbf{Analytical calculation of dispersion relation for 1D acoustic waveguide system.}
To analytically calculate the dispersion relation for 1D acoustic waveguide system in Fig.~\ref{fig1}\textbf{a}, we use the general solution of acoustic pressure for 1D waveguide that is $p_s=A*e^{iks}+B*e^{-iks}$. For a 1D open waveguide, the acoustic particle velocity $u_0$ at one end of the channel ($s=0$) can be represented as a function of pressure $p_0$, $p_L$ at both ends ($s=0$ and $L$),
\begin{equation}
	u_0=-\frac{i}{Z_c}\cot(kL)*p_0+\frac{i}{Z_c}\frac{1}{\sin(kL)}*p_L
\end{equation}
Where $Z_c$ is the characteristic impedance and $k=\omega/v$. For a 1D waveguide with hard boundary condition at one end ($s=L$), the acoustic particle velocity $u_0$ at the other end of the channel ($s=0$) can be represented as a function of pressure $p_0$,

\begin{equation}
	u_0=\frac{i}{Z_c}\tan(kL)*p_0
\end{equation}
At each nodes, the interaction point of $N$ waveguides, the acoustic pressure is constraint by acoustic particle velocity flow conservation law,
\begin{equation}
	\sum_{i=1}^{N}u_i=0
\end{equation}
For the 1D model in Fig.~\ref{fig1}\textbf{a}, there are two nodes which makes,
\begin{equation}
	\left( \begin{array}{cc}
		D_{11}&D_{12}  \\
		D_{21}&D_{22} 
	\end{array}\right).\left( \begin{array}{c}
p_{node1}	\\
	p_{node2}
	\end{array}\right)=0 
\end{equation}
with
\begin{multline}
D_{11}=D_{22}=\tan(k\lambda_0/4)-\cot(k\lambda_0/4)-\cot(k\lambda_0/2)	\\
D_{12}=D_{21}^{*}=\frac{1}{\sin(k\lambda_0/4)}+\frac{e^{-iK_Ba_s}}{\sin(k\lambda_0/2)}
\end{multline}
Where $K_B$ is the Bloch momentum.The dispersion relation of the system can be obtained from the determinant of the matrix. After simplification, it is 

\begin{equation}
\begin{split}
	\cos(K_Ba_s)= & \frac{1}{2}\sin(k\lambda_0/4)\sin(k\lambda_0/2)* \\
	& \lbrace [ \tan(k\lambda_0/4)-\cot(k\lambda_0/4)-\cot(k\lambda_0/2)] ^2 \\
	&-\frac{1}{\sin^2(k\lambda_0/4)}-\frac{1}{\sin^2(k\lambda_0/2)} \rbrace 
\end{split}
\end{equation}

\section*{Acknowledgement}
W.Z. acknowledges support from the Start up Funding from the Ocean University of China and the National Natural Science Foundation of China (Grants No.~12404499). H.-X.S. and Y.G. are supported by the National Natural Science Foundation of China (Grants Nos. 12274183 and 12174159). S.-Q.Y. is supported by the National Key R\&D Program of China (Grant No. 2020YFC1512403). H.X. acknowledges the support of the start-up fund and the direct grant (Grant No. 4053675) from The Chinese University of Hong Kong. B.Z. acknowledges support from Singapore National Research Foundation Competitive Research Program Grant No. NRF-CRP23-2019-0007, Singapore Ministry of Education Academic Research Fund Tier 2 Grant No. MOE-T2EP50123-0007, and Tier 1 Grant No. RG81/23.

\textbf{Author Contributions:} W.Z., H.X. and B.Z. initiated the project. W.Z. perform the numerical simulation. H.-X.S., Y.W., W.Z. and Z.C. fabricated samples. H.-X.S., H.Z., Y.G., Y.W., B.W. and S.-Q.Y. carried out the measurement. W.Z., H.-X.S., H.X. and B.Z. analyzed the results. W.Z., S.-Q.Y., H.-X.S., H.X. and B.Z. wrote the manuscript with input from all authors. B.Z. supervised the project.

\textbf{Competing Interests:} The authors declare no competing interests. 

\textbf{Data availability:} All data are available from the corresponding authors upon reasonable request.

\end{document}